\documentclass[fleqn,usenatbib]{mnras}
\usepackage[T1]{fontenc}
\usepackage{ae,aecompl}

\usepackage{graphicx}
\usepackage{footnote}

\def\spose#1{\hbox to 0pt{#1\hss}}
\newcommand\lsim{\mathrel{\spose{\lower 3.0pt\hbox{$\mathchar"218$}}
     \raise 2.0pt\hbox{$\mathchar"13C$}}}
\newcommand\gsim{\mathrel{\spose{\lower 3.0pt\hbox{$\mathchar"218$}}
     \raise 2.0pt\hbox{$\mathchar"13E$}}}
\newcommand\msun{{\rm \,M_\odot}}

\title[Low-end mass function of the Quintuplet cluster]{Low-end mass function of the Quintuplet cluster}
\author[J. Shin and S. S. Kim]{Jihye Shin$^{1,2}$\thanks{Email: jhshin.jhshin@gmail.com}
and Sungsoo S. Kim$^{3,4}$\thanks{Corresponding author}\\
$^{1}$Kavli Institute for Astronomy and Astrophysics at Peking
University, Yi He Yuan Lu 5, Hai Dian District, Beijing 100871, P.R. China\\
$^{2}$School of Physics, Korea Institute for Advanced Study,
Heogiro 85, Seoul 02455, Republic of Korea\\
$^{3}$Department of Astronomy \& Space Science, Kyung Hee University,
Yongin, Kyungki 17104, Republic of Korea\\
$^{4}$School of Space Research, Kyung Hee University, Yongin, Kyungki
17104, Republic of Korea}

\date{Accepted 1988 December 15. Received 1988 December 14; in original form 1988 October 11}

\pubyear{2016}

\begin{document}
\label{firstpage}
\pagerange{\pageref{firstpage}--\pageref{lastpage}}
\maketitle

\begin{abstract}
The Quintuplet and Arches clusters, which were formed in the harsh
environment of the Galactic Center (GC) a few million years ago,
have been excellent targets for studying the effects of a star-forming
environment on the initial mass function (IMF).
In order to estimate the shape of the low-end IMF of the
Arches cluster, Shin \& Kim devised a novel photometric method that utilizes
pixel intensity histograms (PIHs) of the observed images.  Here, we apply the
PIH method to the Quintuplet cluster and estimate the shape of its low-end IMF
below the magnitude of completeness limit as set by conventional
photometry.  We found that the low-end IMF of the Quintuplet is
consistent with that found for the Arches cluster---Kroupa MF,
with a significant number of
low-mass stars below $1\msun$.  We conclude that the most likely IMFs
of the Quintuplet and the Arches clusters are not too different from the IMFs
found in the Galactic disc. We also find that the observed PIHs and stellar
number density profiles of both clusters are best reproduced when the
clusters are assumed to be at three-dimensional distances of approximately
$100$~pc from the GC.
\end{abstract}


\section{Introduction}

The Quintuplet and Arches are two young ($\sim 2$--4 Myrs old), massive
($\sim 2\times10^4\msun$), and compact ($\leq 1~$pc) star clusters that are
located near ($\sim 30$~pc in projection from) the Galactic center (GC;
\citep{fig99,fig02,kim00,naj04,mar08,esp09}).
Since the GC has unusual star-formation environments such as elevated
temperatures and turbulent velocities in the molecular clouds, strong
magnetic fields, and large tidal forces, these clusters have often been
suspected to have had a shallower initial mass function (IMF) and/or
an elevated low-mass cutoff ($M_l$) in the IMF.  Thus, these clusters have
been regarded as important targets for understanding the effects of the
star-formation environment on the IMFs \citep{mor93}.

Attempts to estimate the IMF of these extraordinary clusters have been carried
out by high-resolution photometries using the infrared (IR) camera onboard
the \textit{Hubble Space Telescope (HST)}, and also using a few ground
IR telescopes with adaptive optics capabilities 
\citep{fig99,sto02,sto05,kim06,esp09,hub12,hab13}.  Yet, the stellar masses
corresponding to the 50-percent completeness limit reach down to only
$\sim 5\msun$ for the Quintuplet and $1.3\msun$ for the Arches
\citep{kim06,hub12}, photometrically resolving stars
fainter than this limit will have to wait until the next-generation IR
space telescopes or extremely large ground telescopes are available.

The current photometric completeness limit for the Arches cluster was overcome
by \citet[Paper I hereafter]{shi15}, who devised a novel photometric method
using the pixel intensity histograms (PIHs) of the observed image.  The
lower-end of the observed PIH contains information on the unresolved,
faint stars.  Paper I compared the
PIHs between the observed images and the artificial images that were
constructed with various synthetic luminosity functions (LFs) for the
cluster.  The synthetic LFs were constructed by combining the observed LF
above the completeness limit and a model LF, which was converted from two
different types of model mass functions (MFs): single power-law MFs and
Kroupa MFs \citep{kro01}. The fore/background stellar population towards the
Arches cluster was built using the Milky Way star-count model of
\citet{wai92}.  They found that the low-end MF for the Arches cluster is
consistent with the IMF found in the Galactic disc, a Kroupa IMF
with a significant number of stars below $1\msun$ that are not resolved by
current IR telescopes.

The Quintuplet cluster is quite similar to the Arches cluster in several
aspects (age, Galactocentric radius in projection, and initial total mass)
except that the current stellar density of the Quintuplet is two orders of
magnitude lower than that of the Arches.  It is conceivable that the two clusters
have been formed with similar initial conditions (total mass and size),
but the Quintuplet, which was formed a few Myrs earlier, has
evolved more dynamically, and thus currently has a lower stellar density \citep{kim00}.
In the present paper, we apply the PIH method to the Quintuplet cluster to see
if its lower-end IMF is also estimated to be similar to that of the Galactic
disk, similar to the Arches.

This paper, a sequel to Paper I, is organized as follows.  In \S 2, we
describe the observational data used in the present study.  We estimate the
fore/background stellar population and background flux in the observed images
towards the Quintuplet cluster in \S 3.  \S 4 is devoted to tracing out the most
plausible low-end MF of the Quintuplet cluster using the PIH method.
A summary and discussion of the results are given in \S 5.
\section{Observational Data and Photometry}

\begin{figure}
\centering
\includegraphics{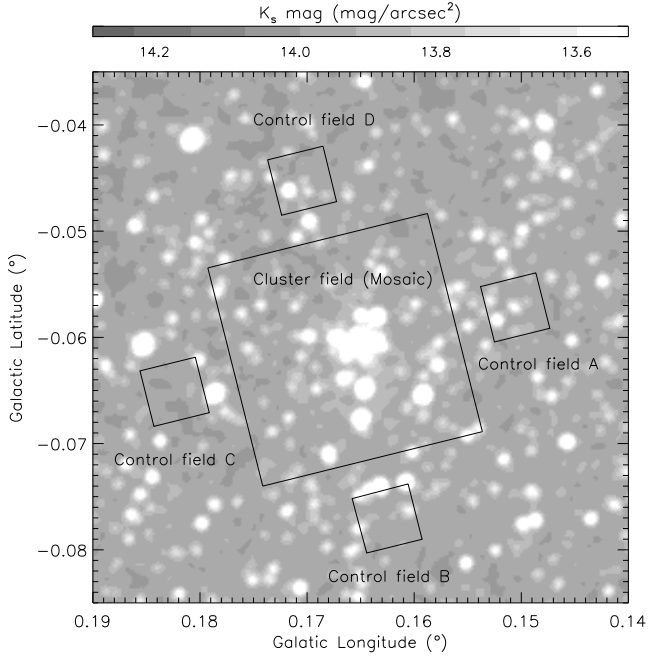}
\caption{Locations and image sizes of the cluster field and the four
control fields used in the present study.  All fields were observed by
the NICMOS2 camera onboard the {\it HST} \citep{fig99}.  The cluster
field was observed in a $4\times4$ mosaic pattern.  The background is
a K$_{\rm s}$-band image from the 2MASS \citep{skr06}.}
\label{location}
\end{figure}

The IR images of the Quintuplet cluster used in this study were obtained
with the NICMOS2 camera onboard the {\it HST} on UT September 13/14, 1997
\citep{fig99}.  The NICMOS2 camera has a field-of-view (FOV) of
$19''.2 \times 19''.2$ and
$256 \times 256$ pixels (each pixel covers $0''.076 \times 0''.076$ of the
sky).  The cluster was observed in a mosaic-pattern of $4 \times 4$,
and four control fields were observed at a distance of $59''$ from the center
of the cluster field (see Fig. \ref{location}).  All image data were taken with
F110W, F160W, and F205W filters, but the present study makes use of only the
F205W images, as they suffer the smallest extinction.  The exposure time
of all F205W images used in the present study is 256~s.

Data were reduced using Space Telescope Institute (STScI) pipeline routines,
and point spread function (PSF) photometry was performed using the DAOFIND
and DAOPHOT packages \citep{ste87} within the Image Reduction and Analysis
Facility (IRAF).\footnote{IRAF is distributed by the National Optical
Astronomy Observatories, which are operated by the Association of Universities
for Research in Astronomy, Inc., under cooperative agreement with the
National Science Foundation.}  The conversion of the PSF flux into the
Vega magnitude was done using the photometric keywords and zero-points
announced on the STScI
webpage\footnote{http://www.stsci.edu/hst/nicmos/performance/photometry}.

Incomplete photometry for faint stars was corrected via recovery fractions; these
are estimated for each 0.5 magnitude bin by planting artificial stars on the
images.  In our analyses in the forthcoming sections, only the portion of
the observed LF whose recovery fraction is higher than 80\% is used.

We adopted a distance modulus of 14.52 for the GC
($R_g=8$~kpc; \citealt{rei93}) and solar-metallicity isochrone at 4 Myr
of the PARSEC model\footnote{http://stev.oapd.inaf.it/cmd} \citep{ble12}
for the conversion between the stellar magnitude and the mass. The
reddening value of each star was estimated from the color excess in the
F160W$-$F205W colors and the extinction law of \citet{nis06} with an
assumption that all stars have the same intrinsic color of
$(\mathrm{F160W}-\mathrm{F205W})_0=-0.05$~mag \citep{fig99,kim06}.
\subsection{The Control Fields}
\begin{figure*}
\centering
\includegraphics{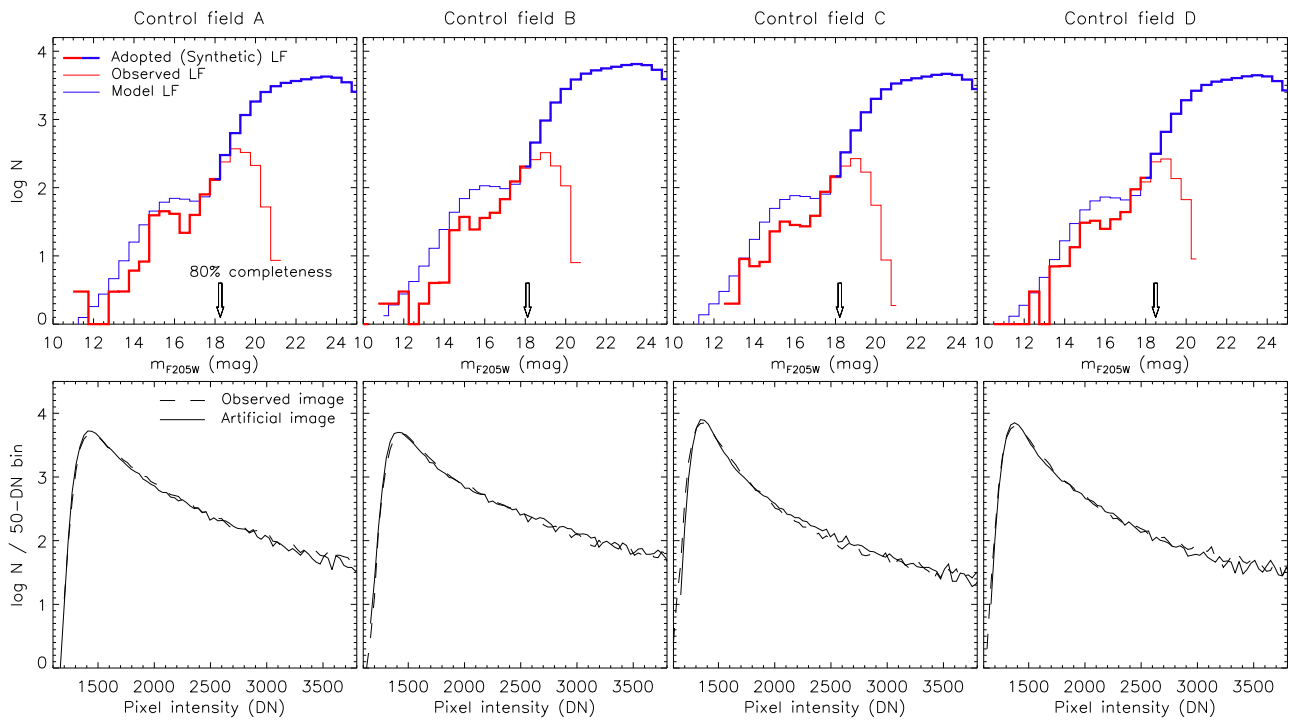}
\caption{Luminosity functions (LFs) and pixel intensity histograms (PIHs)
of the control fields A, B, C, and D. In the upper panels, the red lines are the
completeness-corrected LFs from the observations, and the blue lines are the
model LFs from the star-count model of \citet{wai92}.  We vertically shifted
the model LFs to match them with the observed LFs at $m_{80}$ (arrows), then
built the synthetic LFs (thick lines) by combining the observed LFs for the
bright-part and the model LFs for the faint-part.  The lower panels compare
the observed PIHs (dashed lines) with the PIHs of the artificial images
constructed from the corresponding synthetic LFs (solid lines).  The
histograms are the number of pixels for each 50-DN bin.}
\label{control_hist}
\end{figure*}

Before applying the PIH method to the Quintuplet cluster, we first
build the fore/background stellar LFs and estimate the background flux
in the directions of the control fields.

For the magnitude bins brighter than a magnitude of $80\%$ completeness,
$m_{80}$, of each field (18.3, 18.1, 18.2, and 18.5 mag for control
fields A, B, C, and D, respectively), we adopt the observed,
completeness-corrected F205W LFs from each field.
For the magnitude bins fainter than $m_{80}$, we
construct model LFs using the star-count model of the Milky Way in the K band
of \citet{wai92}.\footnote{We have shown in Paper I that our PIH analyses
for the Arches cluster were not dependent on the choice of the Milky Way
star-count model. We have tried the star-count model of \citet{gir05} and
\citet{pol13} in addition to the model of \citet{wai92} for the Quintuplet cluster
and have reached the same conclusion that the faint-end PIHs constructed
from these models are nearly indistinguishable.} The K magnitudes of the
stellar objects in the star-count
model are reddened with an average extinction of 2.8~mag and observed
standard deviations of 0.07 0.08, 0.06, and 0.08~mag for the control fields A, B, C,
and D, respectively.  Here we assumed that the LFs in the K and F205W bands
are adequately similar.

Since the star-count model may not be accurate towards the GC region due
to a large and spatially varying extinction, the levels of the model LFs from the
star-count model may be slightly different from the observed LFs near $m_{80}$.
We found that the logarithmic model LFs of the control fields A, B, C, and D
are 0.12, 0.26, 0.25, and 0.11 dex higher than the observed LFs at $m_{80}$
respectively, thus we subtract these amounts from our model log LFs.
These offset values are slightly smaller than those found for the
control fields of the Arches cluster in Paper I.

We then construct the final synthetic LF, by merging the bright-part LF from
the PSF photometry and the modified faint-part LF from the star-count model.
The upper panels of Figure \ref{control_hist} show the completeness-corrected
LFs from the observations (red lines), the model LFs modified (shifted vertically
as described above) to match the observed LFs at $m_{80}$ (blue lines), and
the final synthetic LFs for the four control fields (thick lines).

\begin{figure}
\centering
\includegraphics[scale=0.9]{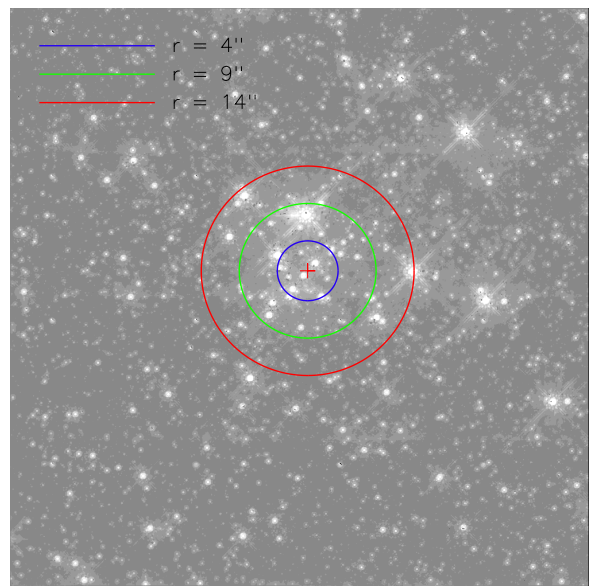}
\caption{Cluster field image obtained with a F205W filter and the three
annuli of 4, 9, and $14''$ (corresponding to 0.16, 0.35, and 0.54~pc, respectively)
from the cluster center, which is marked with a cross.}
\label{annulus}
\end{figure}

Using these synthetic LFs for each control field, we construct artificial
images following the same procedure described in Paper I, which can be
summarized as follows: 1) artificial stars are planted in an empty artificial
image at random locations using a set of 64 model PSFs, 2) the magnitude of
the artificial stars are randomly chosen following the synthetic LF obtained
above, 3) a dark current ($76.8~\mathrm{e}^-$) and readout noise (standard
deviation of $\sqrt{26}~\mathrm{e}^-$) are applied to the artificial image,
and 4) a background flux of 1,225~DN (data number\footnote{The gain,
conversion between the electrons and the data number, of NICMOS2 camera is
$5.4~\mathrm{e}^-$/DN.}) is added to the artificial image.

A background flux is composed of telescopic thermal noise, zodiacal light,
external galactic sources, and cosmic IR background, making it rather
difficult to estimate the background flux theoretically.  As in Paper I,
we estimated the amount of background flux toward the control fields by
comparing our artificial images against the observed images.
We found that adding a background flux of 1,225~DN results in a good
match between the PIHs from the artificial and observed images for all four
of the control fields (see the lower panels of Fig. \ref{control_hist}).
This amount is only slightly larger than that obtained for the control
fields of the Arches cluster in Paper I, 1,190~DN.

The synthetic LFs and the background flux of 1,225~DN obtained for the four
control fields here will also be used for the analyses of the cluster field, given
in \S 4. The 95\% confidence interval from our best-fit background flux
of 1,225 DN is estimated to be $\sim$10 DN. We find that our results
from forthcoming PIH analyses are robust against the uncertainties in
background flux estimation.

\section{The Cluster Field}

\begin{figure*}
\centering
\includegraphics[scale=0.85]{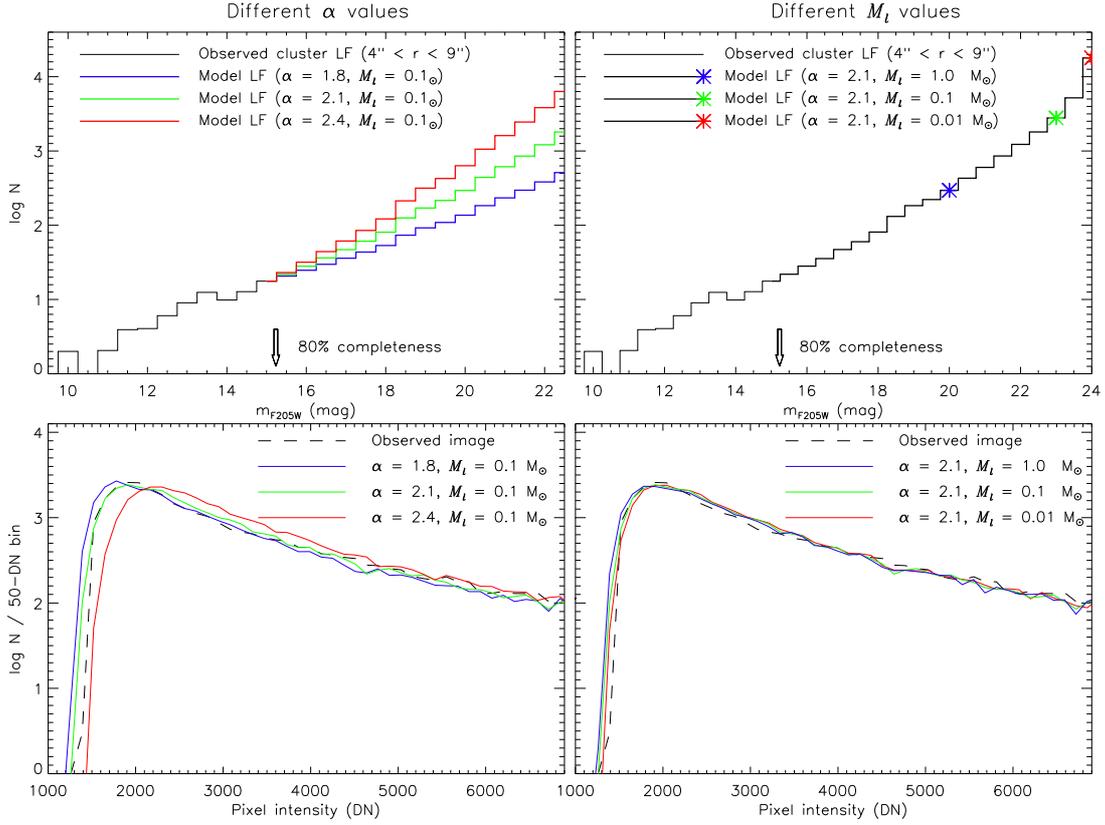}
\caption{LFs and PIHs of the cluster field with different $\alpha$
(left panels) or $M_l$ (right panels) values for the single power-law MFs.
The model LFs, converted from the model MFs, are connected with the observed
cluster LF at $m_{80}$ (upper panels) to form the synthetic LFs.  The PIHs
of the artificial images constructed from the synthetic LFs are compared with
the observed PIHs (lower panels).}
\label{single_diff}
\end{figure*}

\begin{figure*}
\centering
\includegraphics{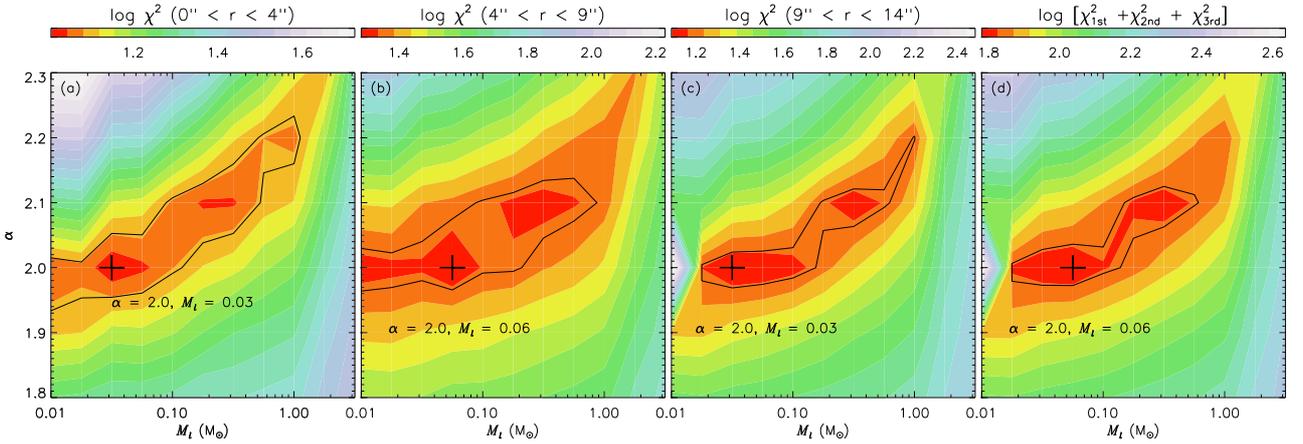}
\caption{Color maps of $\chi^2$ of the cluster field in the $\alpha-M_l$
space for the single power-law MF model.  The first three panels are for
$0''<r<4''$, $4''<r<9''$, and $9''<r<14''$ annuli, respectively, and the
last panel is for the three annuli together.  The minimum $\chi^2$
points are marked with cross symbols. Black contours indicate
the 95\% confidence intervals from the minima.}
\label{single_survey}
\end{figure*}

The stellar number density and LF in the cluster field are highly
position-dependent; the central region of the cluster is confusion limited by
bright stars, while the outskirt of the cluster is limited by the fore/background
stars.  For our analysis here, we use only the central region of the
cluster field and divide it into three annuli of $0''<r<4''$, $4''<r<9''$,
and $9''<r<14''$ (see Fig. \ref{annulus}).  The position of the cluster center
is defined to be the center of 100 brightest stars, R.A. =
$17^{\mathrm{h}}46^{\mathrm{m}}15.24^{\mathrm{s}}$ and dec. =
$-28^{\circ}49'36.12''$. Unlike in Paper I, we make use of the innermost
annulus ($0''<r<4''$), since the stellar number density of the Quintuplet
is two orders of magnitude lower than that of the Arches,
and the undesired contributions of bright stars to the PIHs are much less
significant, even in the innermost region.

Our stellar photometry and completeness test result in $m_{80}=14.4$~mag
for $0''<r<4''$, 15.2~mag for $4''<r<9''$, and 16.8~mag for $9''<r<14''$.
We construct the synthetic LF for the three cluster annuli by merging the
bright-part LF from the PSF photometry and the faint-part LF from models.
The latter consists of 1) the synthetic LF for the fore/background stars,
and 2) the LF converted from a model MF for the cluster stars.  For the
cluster model MF, two models are adopted: 1) single power-law MFs with
various exponents ($\alpha$) and lower mass limits ($M_l$)
\footnote{The function form of the single power-law MF is $dN(M) \propto
M^{-\alpha} dM$ for $M>M_l$, where $dN(M)$ is the number of stars with
masses between $M$ and $M+dM$.}
and 2) dynamically evolved Kroupa MFs \citep{kro01}
\footnote{The Kroupa MF consists of three power-laws,
$\alpha=2.3$ for $M>0.5\msun$, $\alpha=1.3$ for $0.08<M/\msun<0.5$,
and $\alpha=0.3$ for $M<0.08\msun$} with various $M_l$ and
tidal radii ($R_t$).  We try the for a consistency check.

The artificial images for the three cluster annuli are constructed in the
same procedure used for the control fields with a background flux of
1,225~DN, except that the magnitudes of the stellar objects are reddened
with position-dependent extinction values obtained from our extinction map.
The averaged extinction of the cluster field is estimated to be
$3.6\pm0.2$~mag for $0''<r<4''$, $3.4\pm0.4$ mag for $4''<r<9''$, and
$3.0\pm0.2$ mag for $9''<r<14''$ (cf. $3.1\pm0.2$~mag obtained
for $r<12''$ by \citet{fig99} and $3.1\pm0.5$~mag for Wolf-Rayet stars 
in the inner part of the cluster by \citet{lie10}).
Since the area covered by each cluster annulus is rather small,
unlike in the control fields, we used the magnitudes and positions of the
actual observed stars when planting artificial stars brighter than $m_{80}$;
this was done in order to reduce the effects of random choices of magnitudes
and positions of bright stars on the faint-end PIH.

\subsection{Single Power-Law Mass Function}

In this subsection, we describe our analyses performed with the single power-law
MFs with various $\alpha$ and $M_l$ values for the faint-part LF of the cluster.
The upper panels of Figure \ref{single_diff} show a few sample synthetic LFs
for the middle annulus ($4''<r<9''$), whose faint-part cluster LFs are from the
MFs with three different $\alpha$ values (left panel), and three different
$M_l$ values (right panel).  The lower panels of Figure \ref{single_diff}
plot the PIHs of the observed images (dashed lines) and those of the artificial
images constructed from the synthetic LFs in the corresponding upper panels
(solid lines).

The lower panels show that the turnover intensity of the PIH depends on both
the $\alpha$ and $M_l$ values, although it is less sensitive on $M_l$ than $\alpha$.
A few sample experiments shown in these panels already imply that the
best-fit $\alpha$ and $M_l$ for the middle annulus are around 2.1 and
0.1~M$_\odot$, respectively. In order to quantitatively search for the
best-fit $\alpha$ and $M_l$ values, we minimize the $\chi^2$ values between
the PIHs from observed and artificial images. The number of DN bins used
for the $\chi^2$ test is 15.

The color map of $\chi^2$ in the $\alpha$-$M_l$ space for the three cluster
annuli are shown in panels a, b, and c of Figure \ref{single_survey}. The
minimum of $\chi^2$ is near ($\alpha=2.0^{+.03}_{-.03}$, $M_l=0.03^{+.06}_{-.01}
\msun$) for the inner annulus, ($\alpha=2.0^{+.04}_{-.02}$, $M_l=0.06^{+.04}_{-.05}
\msun$) for the middle annulus, and ($\alpha=2.0^{+.02}_{-.02}$,
$M_l=0.03^{+.07}_{-.01} \msun$) for the outer annulus (all the uncertainties
in the present paper are 1-sigma confidence intervals).
The $\chi^2$ contours are elongated along the arc-like curves
connecting the (large $\alpha$, large $M_l$) and (small $\alpha$, small $M_l$)
corners of the $\alpha$-$M_l$ space.  The sum of the three
$\chi^2$ values, which traces the best-fit $\alpha$ and $M_l$ for the three
annuli simultaneously, has a minimum at $\alpha=2.0^{+.03}_{-.03}$ and
$M_l=0.06^{+.05}_{-.02} \msun$ (Fig. \ref{single_survey}d).

Our analyses using single power-law MFs imply that the Quintuplet cluster
has a significant amount of unresolved stars below its current photometric
limit ($\sim 5 \msun$) and that the present-day MF of the Quintuplet is
best reproduced with an $\alpha$ values that is smaller (shallower slope)
than that of the Salpeter IMF.

\subsection{Kroupa Mass Function}

\begin{figure}
\centering
\includegraphics{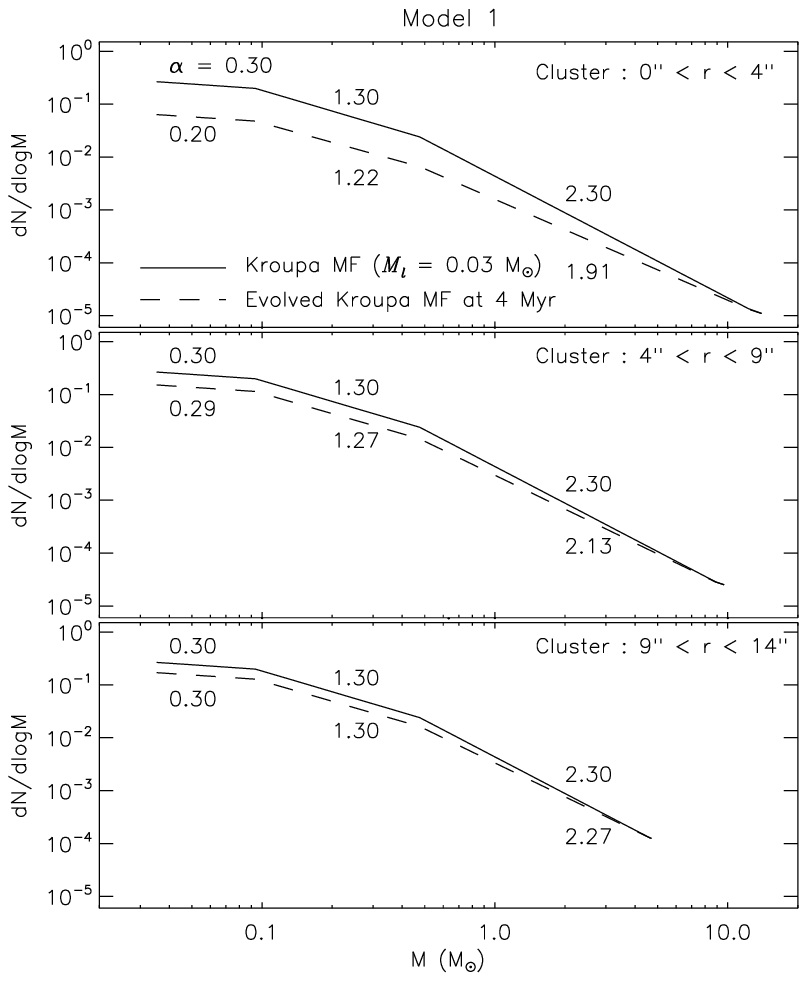}
\caption{Original Kroupa MF (solid lines) and the dynamically evolved Kroupa
MF at 4~Myr calculated using the Fokker-Planck model (dashed lines) below
the masses that correspond to $m_{80}$ for annuli of $0''<r<4''$ (top panel),
$4''<r<9''$ (middle panel), and $9''<r<14''$ (bottom panel).  The heights of
the three MFs are adjusted to match at $m_{80}$.  The original MF has an
$M_l$ of $0.03 \msun$, and the initial tidal radius of the cluster is
1.1 pc (Model 1).}
\label{kroupa_mf}
\end{figure}

\begin{figure}
\centering
\includegraphics{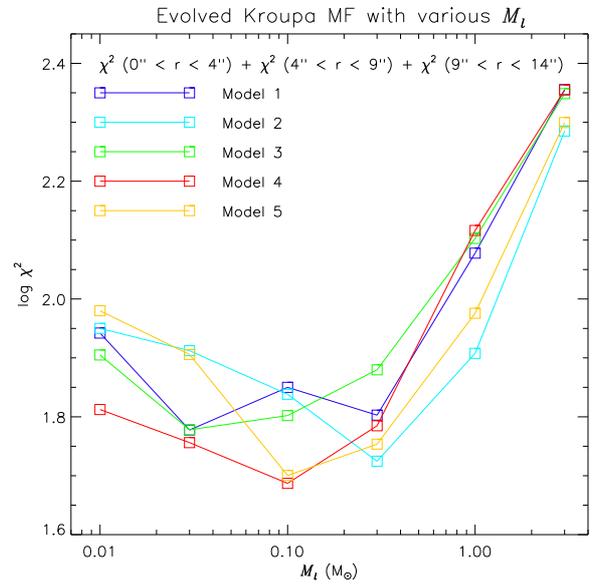}
\caption{$\chi^2$ values summed for all three annuli as functions of $M_l$ for
the evolved Kroupa MFs at 4~Myr. Different colors indicate different $R_t$
models.}
\label{kroupa_survey}
\end{figure}

\begin{figure*}
\centering
\includegraphics{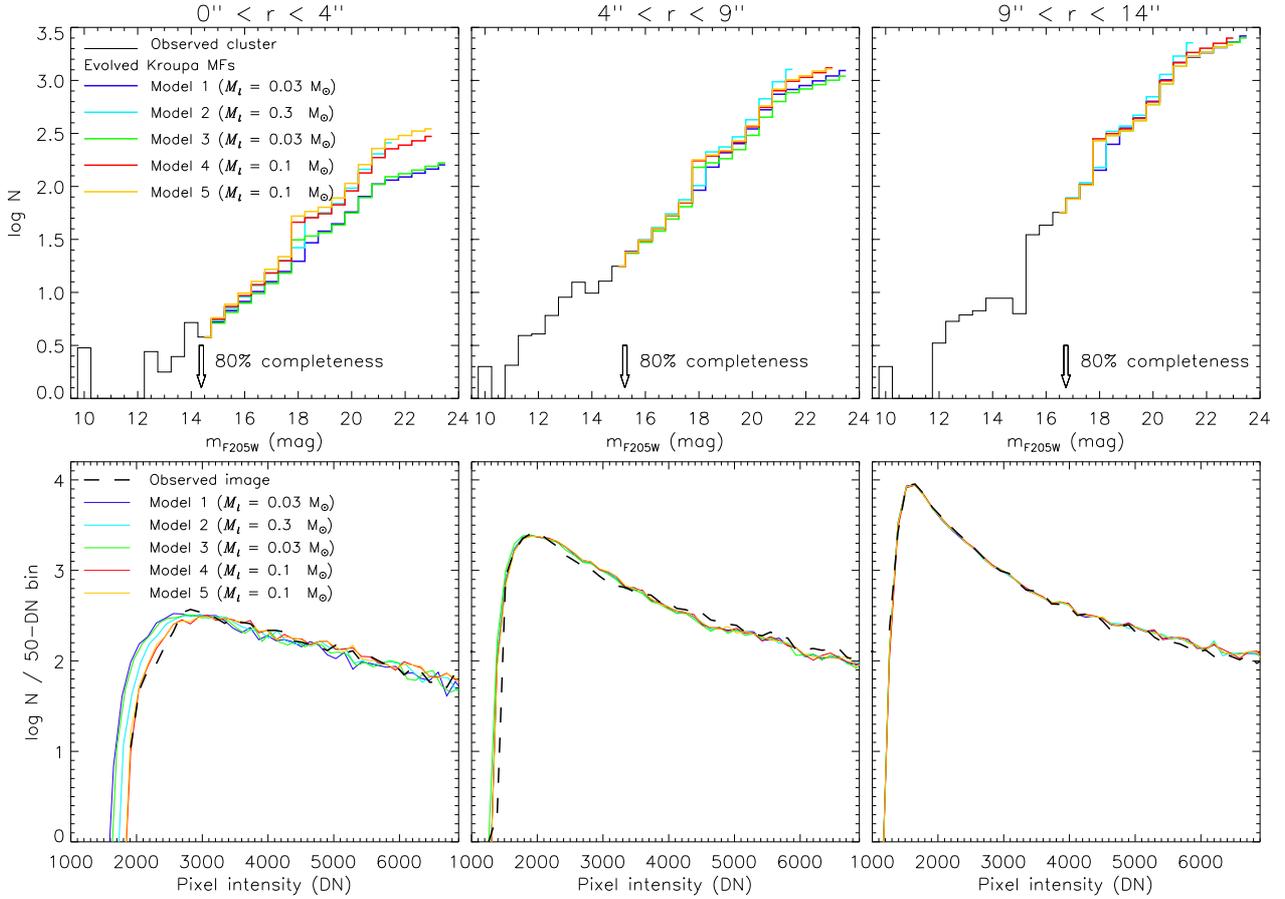}
\caption{LFs and PIHs of the cluster field for the evolved Kroupa MFs with
a best-fit $M_l$ value for each $R_t$ model.  See the caption of Figure 4
for detailed descriptions.}
\label{kroupa_best}
\end{figure*}

\begin{figure}
\centering
\includegraphics{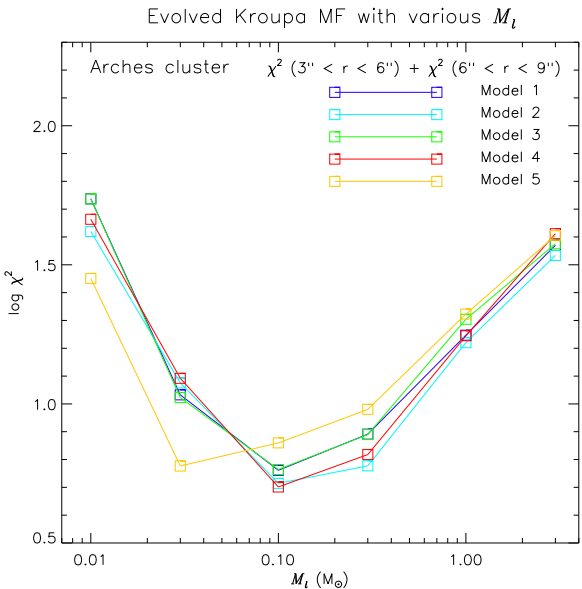}
\caption{$\chi^2$ values as functions of $M_l$ for the evolved Kroupa MFs
at 2~Myr for the {\it Arches cluster}.  $\chi^2$ values are summed for the
annuli of $3''<r<6''$ and $6''<r<9''$.  Different colors indicate different
$R_t$ models.}
\label{arches_survey}
\end{figure}

\begin{figure*}
\centering
\includegraphics{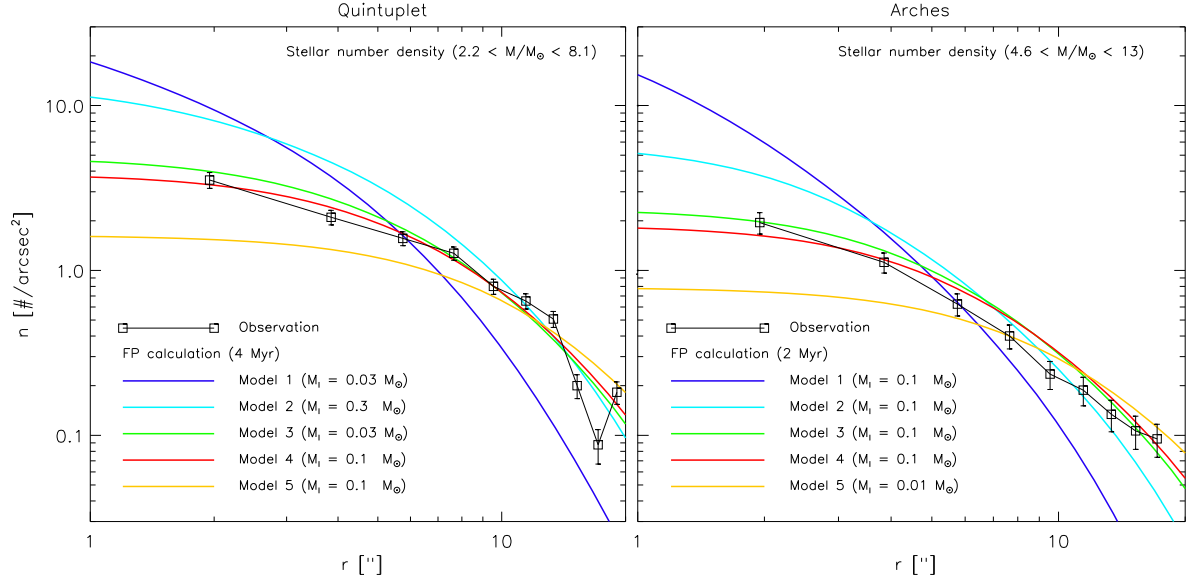}
\caption{Stellar number density profiles of the Quintuplet and
Arches clusters. Black lines are the completeness-corrected density
profiles from the observations, and the colored lines are from our
Fokker-Planck calculations.  Stars within a certain mass range
($2.2 < M/\msun < 8.1$ for the Quintuplet and $4.6 < M/\msun < 13$ for
the Arches) are considered for a comparison.  Error bars represent the
1-$\sigma$ uncertainties.}
\label{den_prof}
\end{figure*}

\begin{table*}
\centering
\caption{Best-fit $M_l$ values for the evolved Kroupa MF \label{tbl-1}}
\begin{tabular}{cccccccccccc}
\hline
Model & $R_t$ & $M_l$ & \multicolumn{4}{c}{$\chi^2$} &  &
    \multicolumn{4}{c}{p-value [\%]} \\ \cline{4-7} \cline{9-12} \\
& [pc] & $[\msun]$ & $0-4''$ & $4-9''$ & $9-14''$ & $0-14''$ & & $0-4''$ &
    $4-9''$ & $9-14''$ & $0-14''$\\
\hline
1 & 1.1 & $0.03^{+.04}_{-.01}$ & 22.56 & 25.04 & 12.32 & 59.91  & &   6.79 & 3.41 & 58.09 & 4.48 \\
2 & 1.7 & $0.30^{+.12}_{-.12}$ & 16.54 & 25.85 &  10.65 &  53.04 & &  28.12 & 2.71 & 71.35 & 14.03 \\
3 & 2.3 & $0.03^{+.10}_{-.01}$ & 14.83 & 27.71 & 17.43  & 59.97 & &  39.00 & 1.55 & 23.39 & 4.43 \\
4 & 2.5 & $0.10^{+.10}_{-.07}$ & 14.41 & 24.16 & 10.08  & 48.64  & & 41.99 & 4.39 & 75.61 & 25.61 \\
5 & 3.5 & $0.10^{+.03}_{-.02}$ & 13.98 & 26.86 & 9.32  &  50.16  & &  45.09 & 2.01 & 81.02 & 21.06\\
\hline
\multicolumn{12}{l}
{\begin{minipage}{12cm}
Note---The $p-$value is the probability of having a $\chi^2$ that
is larger than the value obtained from our $\chi^2$ test between the model
and the observation, whose degree of freedom is 14 for each annulus of
$0''<r<4''$, $4''<r<9''$, and $9''<r<14''$, and 42 for $0''<r<14''$.
\end{minipage}
}
\end{tabular}
\end{table*}

In this subsection, we adopt the Kroupa MFs
\citep{kro01} with various $M_l$ values as a more
realistic initial MF of the cluster.  We allowed them to dynamically evolve
for 4 Myr using the anisotropic Fokker-Planck (FP) method described in
\citet{kim99} and \citet{kim00}.  Following these previous studies on
the dynamics of the star clusters in the GC, we adopted the following initial
conditions for the Quintuplet: density and velocity dispersion structures
from the King model, a total mass of $2\times10^4$~M$_{\odot}$, upper mass
boundary of 150~M$_{\odot}$, and a King concentration parameter of 4.

One of the factors that affect the dynamical evolution of a star cluster
is the tidal radius of the cluster,
\begin{equation}
	R_t = {\left(\frac{M_{cl}}{2M_g}\right)}^{1/3}R_g,
\end{equation}
where $M_{cl}$ is the cluster mass, and $M_g$ is the Galactic mass inside a
Galactocentric radius $R_g$. The $R_g$ of the Quintuplet cluster is
not certain, so we adopt three different values, 30~pc, 100~pc, and 200~pc.
For the Galactic enclosed mass profile $M_g(R_g)$, we adopt two different
models, those by \citet[GT87 hereafter]{gen87} and
\citet[LZM02 hereafter]{lau02}.  We try five different initial
$R_t$ values for the Quintuplet cluster: 1) 1.1~pc
(Model 1; $M_g$ of GT87 at $R_g=30~$pc), 2) 1.7~pc (Model 2; $M_g$ of LZM02
at $R_g=30~$pc), 3) 2.3~pc (Model 3; $M_g$ of GT87 at $R_g=100~$pc), 4)
2.5~pc (Model 4; $M_g$ of LZM02 at $R_g=100~$pc), and 5) 3.5~pc (Model 5;
$M_g$ of GT87 and LZM02 at $R_g=200~$pc; $M_g\sim1.9\times10^9\msun$ at
$R_g=200$~pc for both models).

Figure \ref{kroupa_mf} compares the Kroupa IMF with $M_l=0.03~\msun$ and
$R_t=1.1~$pc (Model 1) and its evolved MF at 4~Myr from our FP calculation.
The MF at 4~Myr is shallower in the inner annulus, while it is steeper in
the outer annulus; this is due to the mass segregation.  Since the degree of mass
segregation may differ for different values of $M_l$ and $R_t$, the adopted
MFs for each annulus depend on the $M_l$ and $R_t$ values of the cluster.

We construct the synthetic LFs and artificial PIHs for different $M_l$ and
$R_t$ values in the same way as in \S 4.1. Figure \ref{kroupa_survey} shows
the logarithmic profiles of the summed $\chi^2$ of the three annuli as a function
of $M_l$ for five different initial $R_t$ values.
 The minimum $\chi^2$ value is found at $M_l=0.03^{+.04}_{-.01}\msun$
for $R_t=1.1$~pc (Model 1),
$M_l=0.30^{+.12}_{-.12}\msun$ for $R_t=1.7$~pc (Model 2),
$M_l=0.03^{+.10}_{-.01}\msun$ for $R_t=2.3$~pc (Model 3),
$M_l=0.10^{+.10}_{-.07}\msun$ for $R_t=2.5$~pc (Model 4),
and $M_l=0.10^{+.03}_{-.02}\msun$ for $R_t=3.5$~pc (Model 5). 
The best-fit $M_l$ values are
found to be near or smaller than $0.3\msun$ for a wide range of $R_t$ models.
Thus, it is almost certain that a significant number of low-mass stars below the
current photometric limit ($5\msun$) of the cluster is required to match the observed
PIH of the Quintuplet cluster.  This is consistent with the findings of Paper I for
the Arches cluster (only one $R_t$ value was tried for the Arches cluster in
Paper I, however). The $\chi^2$ values and their p-values (statistical
significance) for the best-fit parameter set of each $R_t$ model are presented
in Table \ref{tbl-1}. The p-value for the entire annuli ($0''<r<14''$) is acceptably
high only for Model 4 and 5.

The synthetic LFs and PIHs constructed with the best-fit $M_l$ value for each
$R_t$ model are shown in Figure \ref{kroupa_best}.  The sensitivity of
the artificial PIH on $R_t$ model is the largest in the inner annulus.
The observed PIH in the inner annulus is best reproduced by the artificial
PIHs of $R_t$ Models 4 and 5 with $M_l=0.1\msun$. This can also be seen in
Figure \ref{kroupa_survey}, where Models 4 and 5
have the two smallest $\chi^2$ values among all the data points.
This indicates that our analyses prefers the $R_t$ Models 4 and 5, which
experiences the least amount of mass segregation due to their larger $R_t$
values (Models 4 and 5 have steeper LFs in the inner annulus than the other
models; see the upper-left panel of Fig. \ref{kroupa_best}).

In case of the Arches cluster, Paper I adopted only one initial value of $R_t$,
1.1~pc, because the PIH of the best-fit IMF obtained with that $R_t$ value
had a very good match with the observed PIH (the average $\chi^2$ value for
the two annuli was only $\sim0.7$).  In case of the Quintuplet cluster, our
initial trial for the $R_t$ value, 1.1~pc, did not yield such small averages of
$\chi^2$ values.  This is why we tried five different $R_t$ values for the
Quintuplet in the present study. We found that the initial value of $R_t$ for the
Quintuplet needs to be near or larger than 2.5~pc in order to result in as
good a match as seen for the Arches.

The main cause of this difference between the Arches and the Quintuplet
is the degree of dynamical evolution (or the age).  The Quintuplet cluster
has undergone greater mass segregation, thus the dependence of the LF on
$R_t$ is stronger in the Quintuplet.  To check if this is indeed the case,
we revisit the Arches cluster and perform the same analyses that we do in
Paper I, i.e., for five different initial $R_t$ values of 1.1, 1.7, 2.3, 2.5,
and 3.5 pc.  Figure \ref{arches_survey} plots the $R_t$ dependence of
the total $\chi^2$ (summed for the two annuli) as a function of $M_l$ for the
Arches cluster, and shows that the total $\chi^2$ curve varies only slightly for
different $R_t$ values.

In order to see if some of our $R_t$ models can be excluded,
we compared the stellar number density profiles from observations to our
FP calculations of model clusters with a Kroupa MF and the best-fit $M_l$
value for each $R_t$ model (see Fig. \ref{den_prof}).
For the number density profiles, we consider stellar mass ranges of
$2.2<M/\msun<8.1$ and $4.6<M/\msun<13$, for the Quintuplet and Arches
clusters, respectively.  These mass ranges correspond to $\pm~1$~mag from
the magnitudes of $50\%$ completeness at $r\sim2''$, which are 16.8~mag
and 15.9~mag. Here, the observed densities are the ones that are
completeness-corrected by the estimated recovery fractions.

For both the Quintuplet and Arches clusters, the projected density profiles of
$R_t$ Models 1, 2, and 5 show significant deviations from the observations
at the inner region ($r\lsim5''$), while those of Models 3 and 4 reproduce the
reasonably well the observation for a wide radial range. Thus, the models with
an initial $R_t$ value smaller than 2 pc ($R_g$ smaller than $\sim 30$~pc) or
larger than 3.5 pc ($R_g$ larger than 200~pc) can be excluded. The
$R_t$ range of $2-3.5$pc for these cluster is consistent with a recent study
on the extended structure of the Arches cluster \citep{hos15}, which found
that the cluster does not exhibit any King-like tidal radius out to a radius of
at least 2.8~pc. Among our five initial $R_t$ models, only Model 4 agrees
with the observed PIH
and number density profile simultaneously, implying that both clusters
are probably located approximately 100~pc away from the GC.  The fact that
both clusters have similar $R_g$ estimates strengthens the notion that
the two clusters were formed in the same or neighboring star-forming
region(s) with similar initial masses, but at different epochs.

We also find that the evolved Kroupa MFs well reproduce the observed
top-heavy MFs in the central regions of the Quintuplet and Arches clusters.
\citet{hub12} estimate the MF slope at the center ($r<0.5$pc) of the Quintuplet
to be $\alpha=1.68$ for a mass range of $5<M/\msun<40$, whereas our
Model 4 has $\alpha=1.70\pm{0.08}$ for the same radial and mass range. The
central MF slope for the Arches cluster is found to be 1.5-1.88
\citep{hab13,esp09}, while our Model 4 gives $\alpha=1.85\pm{0.06}$ for a mass
range of $M>10\msun$ and radial range of $r<0.2$pc.

\section{Summary \& Discussion}

\begin{figure}
\centering
\includegraphics{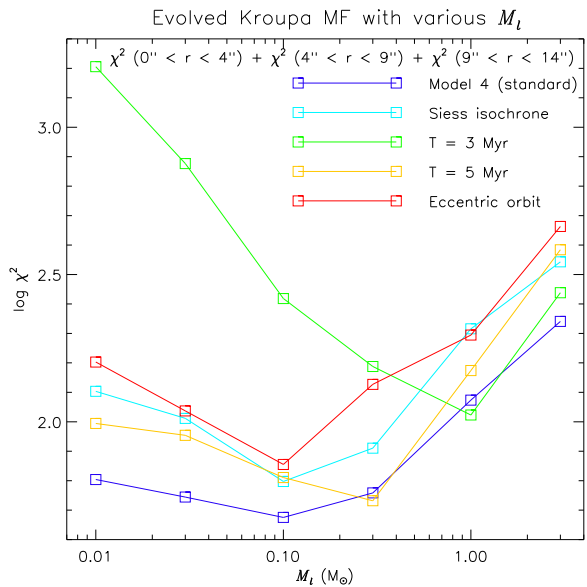}
\caption{Comparison of $\chi^2$ curves between our $R_t$ model 4 (blue)
and its variations (sky blue for the Siess isochrone model, yellow green
and yellow for cluster ages of 3~Myr and 5~Myr, respectively, and red for
the eccentric orbit).}
\label{compare}
\end{figure}

Using the PIH method developed in Paper I, we estimated the
shape of the low-end MF for the Quintuplet cluster, which is a young,
compact star cluster located in the GC.  We have adopted two model MFs
for our PIH analyses, a single power-law MF and the Kroupa MF.

In the case of a single power-law MF, which was tried for a consistency
check, the best-fits between the observed
and model PIHs for $r<14''$ were found at $\alpha=2.0\pm0.03$ and
$M_l=0.06^{+.05}_{-.02}\msun$.  This MF has a shallower $\alpha$
value than that of the Salpeter IMF, $\alpha = 2.35$, and has a lower $M_l$
value than the current 50~\% completeness limit of conventional photometry.

For the Kroupa MF model, we have dynamically evolved the MFs using the
Fokker-Planck methods and compared them with the observed
PIH.  The best-fits between the observed and model PIHs for $r<14''$
were found at $M_l=0.03^{+.04}_{-.01}\msun$
for $R_t=1.1$~pc, $M_l=0.30^{+.12}_{-.12}\msun$ for $R_t=1.7$~pc,
$M_l=0.03^{+.10}_{-.01}\msun$ for $R_t=2.3$~pc,
$M_l=0.10^{+.10}_{-.07}\msun$ for $R_t=2.5$~pc,
and $M_l=0.10^{+.03}_{-.02}\msun$ for $R_t=3.5$~pc.

We found that
the observed PIH of the Quintuplet is modeled reasonably well with the evolved
Kroupa MF, and that the cluster has a significant number of faint stars that
are not resolved by current IR telescopes.  This implies that the IMF of the
Quintuplet is not substantially different from the
IMFs found in the Galactic disc, as is the case for the Arches cluster.

We also found that the observed PIHs and stellar number density profiles
of both the Quintuplet and Arches clusters are best reproduced when the clusters
are assumed to be at three-dimensional distances of $\sim 100$~pc from the GC.
The fact that both clusters have similar $R_g$ estimates strengthens the
notion that the two clusters were formed in the same or neighboring
star-forming region(s) with similar initial masses but at different
epochs.

In order to check if our PIH analyses are dependent on the choice
of isochrone, we have tried the Siess isochrones \citep{sie00} in addition
to the PARSEC model.  Figure \ref{compare} compares the total $\chi^2$ values
of Model 4 for PARSEC with Siess isochrone models. The Siess model yields
generally higher total $\chi^2$ values than the PARSEC model, but results
in the minimum $\chi^2$ value at the same $M_l$ as the PARSEC model.
In case of a single power-law MF, we find that our PIH analysis
with the Siess model results in the $\chi^2$ distribution very similar to
that shown in Figure \ref{single_survey} except that the minimum $\chi^2$
values are found at $M_l=0.06^{+.05}_{-.02} \msun$ for the inner and outer annuli
instead of at $0.03^{+.07}_{-.01} \msun$.
Thus the locations of minimum $\chi^2$ values are not sensitive to the
choice of the isochrone.

The age of the Quintuplet is estimated to be 3-5~Myr \citep{fig99}.
Inaccurate estimation of the cluster age can affect not only the
mass-luminosity relation of the stars but also the degree of dynamical
evolution (mass segregation and selective evaporation of lighter stars).
In order to check the effect
of cluster age on our results, we have tried cluster ages of 3 and 5~Myr
for Model 4 in addition to 4~Myr, which was assumed throughout the paper.
Figure \ref{compare} compares the total $\chi^2$ values of Model 4 for
cluster ages, 3, 4, and 5~Myr. The $\chi^2$ profile of 5~Myr is quite
similar to that of 4~Myr, but its minimum $\chi^2$ is slightly larger
than that of 4~Myr. The $\chi^2$ profile of 3~Myr is significantly different
from that of 4~Myr, and its minimum $\chi^2$ is located at 10 times larger
$M_l$ than that of 4~Myr. Howerver, the minimum $\chi^2$ value is
considerably higher than those of the 4 and 5~Myr cases, implying that
our PIH analysis is much less consistent with a cluster age of 3~Myr.

Studies on the proper motion of the Quintuplet \citep{sto14} suggest
that the cluster's orbit is probably elliptical rather than circular, whereas
circular orbits are assumed for our FP calculations throughout the paper.
The MF at a given radial bin in the cluster may be affected by the
evolution of the tidal radius, which is determined by the orbit of the cluster.
To see the impact of the non-circular orbit on our PIH analyses, we have
tried an eccentric orbit of \citet{sto14} for Model 4, whose current line-of-sight
distance from the GC is 100 pc. Figure \ref{compare} shows that the shape
of the $\chi^2$ profile for the eccentric orbit is very similar to that for the
circular orbit and that the $\chi^2$ values for the eccentric orbit are larger
at all $M_l$ values. This demonstrates that the exact shape of the cluster
orbit does not significantly affect our PIH analyses.

\section*{Acknowledgments}

We thank Pavel Kroupa for helpful discussion and comments.
S.S.K. was supported by the National Research Foundation
grant funded by the Ministry of Science, ICT and Future Planning of Korea
(NRF-2014R1A2A1A11052367).
J.S.'s work was supported by the International Cooperation and Exchange
Program (No. 11450110400) funded by the National Natural Science Foundation
of China (NSFC).


\end{document}